# Effect of shatterproof polymer film application on the fracture types and strength of glass subject to bending load

**Tomohisa Kojima, Ryohei Momokawa, Takuma Matsuo, Mitsuo Notomi**

**Highlights:**

- Three-point bending tests and fracture surface observations were conducted on float glass with a shatterproof polymer film applied.

- By varying the support span of the specimen, failure mode transitioned from bending failure to Hertzian fracture.

- The fractured area subjected to Hertzian fracture was larger with film application.

**Effect of shatterproof polymer film application on the fracture types and strength of glass subject to bending load**

Author names and affiliations

**Tomohisa Kojima[1]  (corresponding author)**

Department of Mechanical Engineering, School of Science and Technology, Meiji University

1-1-1, Higashimita, Tama-ku, Kawasaki-shi, Kanagawa, 214-8571, Japan

E-mail: kojima.31k@g.chuo-u.ac.jp

**Ryohei Momokawa**

Department of Mechanical Engineering, Graduate School of Science and Technology, Meiji University

1-1-1, Higashimita, Tama-ku, Kawasaki-shi, Kanagawa, 214-8571, Japan

**Takuma Matsuo**

Department of Mechanical Engineering, School of Science and Technology, Meiji University

1-1-1, Higashimita, Tama-ku, Kawasaki-shi, Kanagawa, 214-8571, Japan

**Mitsuo Notomi**

Department of Mechanical Engineering, School of Science and Technology, Meiji University

1-1-1, Higashimita, Tama-ku, Kawasaki-shi, Kanagawa, 214-8571, Japan

**Abstract**

Shatterproof polymer films are widely for windows used because they can be easily installed on existing glass windows to improve safety. Applying them to glass plates has been reported to not only prevent fragments from scattering but also increase load-bearing capacity and penetration resistance. However, the clarification of their mechanism and quantitative evaluation are still insufficient because the effect of film application on the strength and failure mode of glass under quasi-static loading has not been investigated. In this study, three-point bending tests and fracture surface observations were conducted on a float glass with a shatterproof polymer film. The stress field formed inside the glass was visualised during the tests using the photoelastic method. By varying the support span of the specimen, the deformation mode was varied to generate three types of failures: bending, shear caused by Hertzian contact stress, and mixed-mode failures. Under the conditions in the present study, the breaking loads of the specimens with and without film were almost the same; however, the fracture surface observation indicated that the area subjected to shear failure caused by Hertzian contact stress was larger with film application. Finally, the effect of the film thickness on the breaking load due to bending deformation was theoretically predicted.

---

[1] Present address: Department of Precision Mechanics, Faculty of Science and Engineering, Chuo University
 1-13-27, Kasuga, Bunkyo-ku, Tokyo, 112-8551, Japan

1 Introduction

Glass is widely used as a plate material for structures and automobile windows because of its high-light transmission. Although glass has a theoretical strength classified in the strongest group among various materials, it is known to be a brittle material with low practical strength owing to the presence of many surface micro-flaws called Griffith flaws [1]. Therefore, increasing the strength of glass is critical for improving its safety and security.

In recent years, many studies have been conducted on the response of glass plates under impact loading [2–4]. Much research has also been conducted on the response to high-speed loading such as blasting [5–7]. Many studies have focused on laminated glass, which has a polymer layer sandwiched between glass plates [8–11]. Several researchers have built and analysed finite element method (FEM) models that include the failure modes of laminated glass and proposed methods to prevent failure [12–15]. Ge et al. [6] investigated fragment projection velocity for architectural glass panels that fail under blast loads to predict fragment flight paths. They developed empirical equations for glass fragment flight trajectories by combining theory and explosion testing. Hidallana-Gamage et al. [7,13] built an FEM model of laminated glass panels that fail under blast loading and examined their design criteria. Binar et al. [12] built an FEM model for each layer of multi-layered transparent armour glass to reproduce the fracture response due to the high-velocity impact of a bullet. Müller-Braun et al. [8] experimentally investigated the differences in strength, and thus the distribution of microdefects, in various areas of a laminated windscreen. Considering the curved geometry of the windscreen, the laminate was separated, and the glass surface facing the interlayer was tested for fracture. They statistically evaluated the results of destructive strength tests and reported that glass surfaces facing the interlayer showed relatively high strength, while the screen-printed areas with enamel showed lower strength, independent of the interlayer. They also reported slightly higher strength on concave surfaces than on convex surfaces. Wang et al. [15]

developed a cohesive element-based FEM model for a tempered laminated windshield used in high-speed trains considering residual stresses to reproduce the distribution of cracks associated with fracture. They used the built FEM model to investigate the effect of the glass and polymer layer thicknesses on the energy absorption performance. Further, several studies have investigated how the use of different polymeric materials in the interlayer of laminated glass alters its mechanical response [16,17].

Glass plates applied with a shatterproof polymer film were the subject of this study. Similar to laminated glass, this is a composite plate consisting of a glass plate and a polymer layer, but it differs from laminated glass in the following two points: 1) laminated glass has a three-layer structure with a polymer layer between two glass plates, whereas a glass plate with a shatterproof film has a two-layer structure consisting of a glass plate and a polymer film. 2) In a glass plate with a shatterproof film, the polymer layer (film) is often even thinner than that in a laminated glass, which is several-tenths of the thickness of the glass plate. Therefore, quantitative evaluation of the effect of the film on failure can be challenging.

Shatterproof polymer films are used to improve the safety of glass plates. They can be easily installed on existing glass windows to prevent secondary damage from glass breakage due to disasters or accidents or increase the safety of window glass in aging facilities. Applying a shatterproof polymer film to glass plates has been reported to not only prevent fragments from scattering but also increase load-bearing capacity and penetration resistance. Dietz [18] reported a qualitative improvement in impact resistance by applying a polymer film to the side glass of an automobile. Van Dam et al. [19] conducted small-scale drop weight tests using polymer film-applied glass plates and studied in detail how the velocity of the impactor was attenuated. Kojima et al. [20,21] quantitatively evaluated the improvement in impact strength when a polymer film was attached to a glass plate by a puncture impact test using a drop weight. In addition, the failure modes of glass plates with polymer films were investigated using high-speed imaging with the shadowgraph method; the glass plates broke in a mixed mode of bending fracture owing to deflection and Hertzian fracture owing to contact with the impactor [22,23]. These studies proved that the impact resistance of glass plates can be easily improved by applying a polymer film. However, the clarification of its mechanism and quantitative evaluation are still insufficient because the effects of film application on the strength and failure mode of glass with a film under quasi-static loading have not been investigated.

To determine the strength of a glass plate, it is necessary to examine the influence of the failure mode of the glass plate. Glass plates that fail due to impact or contact loading are known to fracture based on two different failure modes, namely, bending failure and Hertzian fracture [24]. Bending failure occurs when the loading object is large; thus, the loading area is large and is caused by the tensile stress acting on the surface behind the loading surface by bending deformation. Additionally, the range in which breakage occurs is wide. A Hertzian fracture occurs when the loading object is small; thus, the loading area is small and is caused by tensile stress from the stress concentration on the loading surface. This failure mode is characterised by a small fracture area and a conical fragment of the Hertzian cone.

These failure modes are commonly observed in window glass fractures. Hertzian fracture is also observed in indentation tests on ceramics and is used to evaluate strength properties such as the fracture toughness of glass by indentation tests [25,26]. This has also been observed and investigated in several ballistic studies [27]. However, few studies have investigated these failure modes simultaneously. It is important to identify these failure modes in the failure analysis. Previous studies reported that glass plates subjected to low-velocity impacts fail in a mixed mode of bending failure and Hertzian fracture. It has also been reported that reproducing the failure modes at the onset of failure using FEM is difficult under low-velocity impact loading [28].

This clarifies the effect of polymer film application on the failure mode and strength of a glass plate subjected to quasi-static bending and indentation loading. A three-point bending test and fracture surface observations were conducted. The stress field formed inside the glass was visualised using the photoelastic method during the tests. The effect of the film application was examined by comparing the test results with the bending theory of a beam.

## 2 Methods

### 2.1 Theory for bending failure of single and layered beams

The critical load for a glass with a shatterproof film at which bending failure occurs can be obtained by considering it as a composite beam. Figure 1 shows a schematic of the three-point bending test according to ISO14704 [29] (Fig. 1(a)) and the bending deformation of a composite beam of glass with a shatterproof film applied (Fig. 1(b)). Using a single glass specimen without a shatterproof film, the tensile strength $\sigma_c$ can be obtained from the following equation.

$$\sigma_c = \frac{3 P_c L}{2 w t_g^2}, \quad (1)$$

where $P_c$ is the breaking load, $L$ is the support span, $w$ is the width of the specimen, and $t_g$ is the thickness of the single-layer specimen.

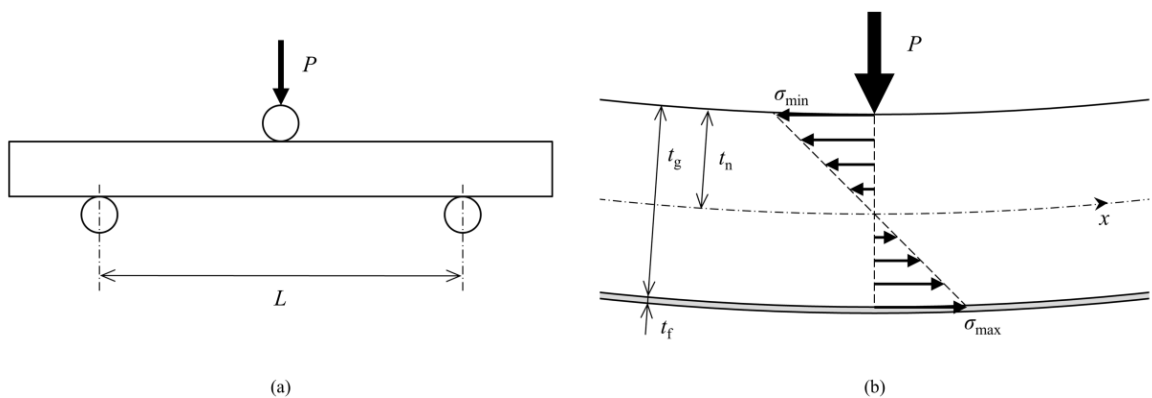

**Fig. 1** Schematic of (a) the three-point bending test and (b) the bending deformation of a composite beam of glass with the shatterproof film applied

Using the theory of a composite beam, the neutral axis of the bending deformation of glass with film can be expressed by the following equation [30].

$$t_n = \frac{A_g \frac{t_g}{2} + A_f \left(\frac{E_f}{E_g}\right)\left(t_g + \frac{t_f}{2}\right)}{A_g + \left(\frac{E_f}{E_g}\right) A_f} = \frac{\frac{t_g^2}{2} + \left(\frac{E_f}{E_g}\right) t_f \left(t_g + \frac{t_f}{2}\right)}{t_g + \left(\frac{E_f}{E_g}\right) t_f}, \quad (2)$$

where $A$ is the cross-sectional area, $E$ is the Young's modulus, $t$ is the thickness, and the subscripts g and f denote glass and film, respectively. The maximum tensile stress $\sigma_{g,\,max}$ occurring in the glass on the surface to which the film is applied is expressed by the following equation.

$$\sigma_{g,max} = \frac{M(t_g - t_n)}{I_g + \left(\frac{E_f}{E_g}\right) I_f}, \quad (3)$$

where $M$ is the bending moment and $I$ is the second moment of area, each of which is expressed by the following equations.

$$M = \frac{1}{4}PL, \tag{4}$$

$$\begin{cases} I_g = \dfrac{wt_g^{\,3}}{12} + wt_g\left(t_n - \dfrac{t_g}{2}\right)^2 \\ I_f = \dfrac{wt_f^{\,3}}{12} + wt_f\left(t_g + \dfrac{t_f}{2} - t_n\right)^2 \end{cases}, \tag{5}$$

where $P$ is the load. From Eqs. (3)–(5), the breaking load of the glass with the film is expressed as follows.

$$P_{c,\text{ with film}} = \frac{4\left[I_g + \left(\dfrac{E_f}{E_g}\right)I_f\right]}{L(t_g - t_n)}\sigma_c \tag{6}$$

Therefore, from Eqs. (1) and (6), the ratio of the breaking load of the glass with and without film can be expressed as a function of the cross-sectional dimensions and Young's modulus of the glass and film in Eq. (7).

$$P_{c,\text{ with film}} / P_{c,\text{ without film}} = \frac{6\left[I_g + \left(\dfrac{E_f}{E_g}\right)I_f\right]}{(t_g - t_n)t_g^{\,2}w} \tag{7}$$

2.2 Experimental method

    The specimens were cut from a 6-mm-thick float glass plate and optically polished on the sides to visualise the stress field using the photoelastic method (Fig. 2). A polystyrene shatterproof film (ULTRA S600, 3M Japan Co., Ltd.) was applied to the specimen. Three-point bending tests were conducted using a universal testing machine (AGS-H, Shimadzu Corporation) according to the method

specified in ISO 14704. The tests were performed under displacement control. The specimens were set such that the film-applied surface was opposite the surface with the loading point. Tables 1 and 2 list the material parameters and experimental conditions, respectively. The support span was varied as the parameter. Three tests were performed for each condition. Considering the random nature of glass strength, this is a small number of tests. However, since there was a small variation in the test results, which will be discussed in the latter section, we determined that three tests were sufficient in the present study.

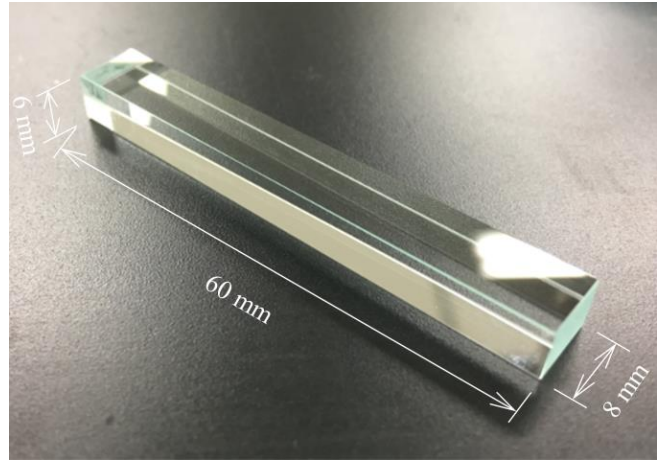

**Fig. 2** Test piece

**Table 1** Material parameters

| Young's modulus | | Thickness | |
|---|---|---|---|
| Glass | Film | Glass | Film |
| 71.3 GPa | 3.7 GPa | 6 mm | 0.19 mm |

**Table 2** Test conditions

| Punch and support pin radius | Support span $L$ | Test speed |
|---|---|---|
| 2 mm | 10, 20, 40 mm | 0.5 mm/min |

The stress field formed inside the specimen during the test was visualised by photoelasticity. A schematic of the photoelastic experimental setup used in this study is presented in Fig. 3. After the test, the fracture surfaces of the specimen were observed using an optical microscope (ECLIPSE LV150, Nikon). Before observation, gold was deposited on the fracture surface using an auto-fine coater (JFC-1100E, JEOL Ltd.) to facilitate observation.

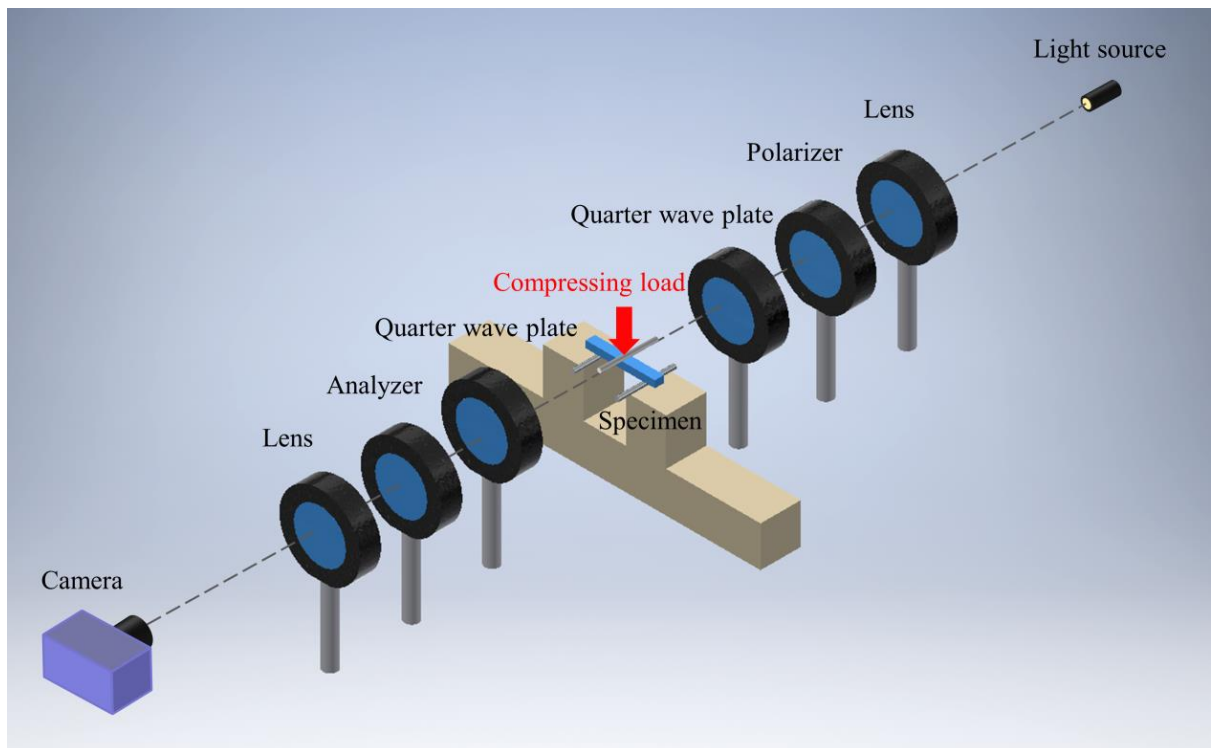

**Fig. 3** Schematic of the photoelastic experimental setup

## 3 Results

### 3.1 Load–displacement relationship

Figure 4 shows the load–displacement curves obtained from the tests (representative data). The variability of the experimental results was small; for example, the standard deviation of the breaking load of the glass without film was less than 10%. The displacement of the specimen with the film was slightly larger than that of the specimen without the film under the same load. This may be due to the compressive deformation of the film pressed by the supports. In conditions with the same support span, the breaking load was approximately the same regardless of whether the film was applied.

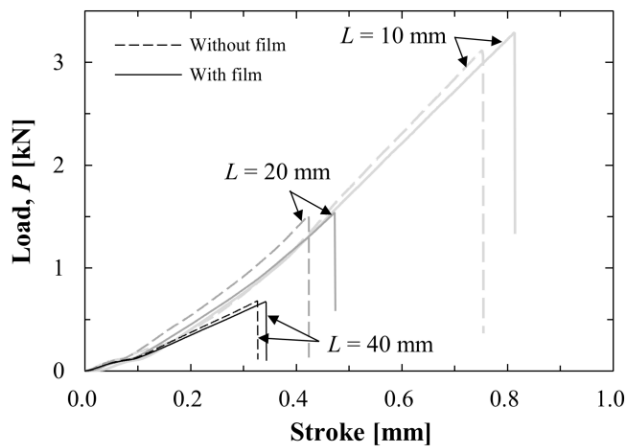

**Fig. 4** Load–displacement curves (representative data)

### 3.2 Photoelastic image

Figures 5–7 show the photoelastic images of the specimens immediately before fracture. The brightness of the images was increased to improve visibility. The photoelastic images obtained were almost the same under the same conditions. Therefore, only representative data are shown in Figs. 5–7. At $L = 40$ mm, a fringe pattern extended in the longitudinal direction of the specimen between the punch and the supports (Fig. 5). As the support span decreased, the longitudinal fringe decreased, and the fringe extending obliquely upward from the supports increased (Figs. 6 and 7). These photoelastic images indicate that the stress field formed inside the specimen changed with the support span. The fringe pattern extending in the longitudinal direction of the specimen is considered to be the stress field formed by the bending deformation of the specimen, whereas the fringe patterns extending obliquely upward from the supports are considered to be the stress field formed by the contact between the punch, supports, and specimen. Thus, by changing the support span, the type of deformation of the specimen changed, resulting in the formation of two types of stress fields in the specimen: one owing to bending and the other owing to contact.

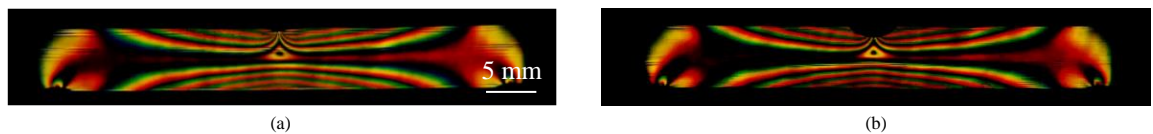

(a)                                      (b)

**Fig. 5** Photoelastic isochromatic map of the specimen just before fracture ($L = 40$ mm); (a) without film, (b) with film

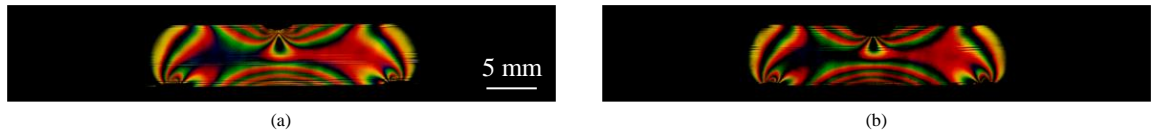

**Fig. 6** Photoelastic isochromatic map of the specimen just before fracture ($L$ = 20 mm); (a) without film, (b) with film

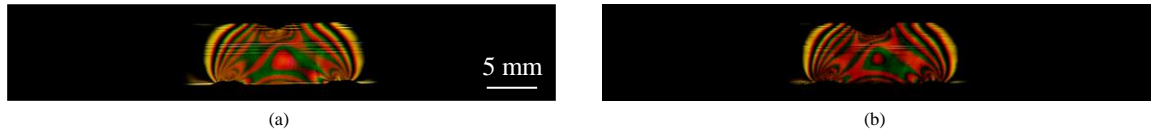

**Fig. 7** Photoelastic isochromatic map of the specimen just before fracture ($L$ = 10 mm); (a) without film, (b) with film

**3.3** Fracture patterns

Figures 8–10 show the vicinity of the fracture location of the specimens broken at support spans of $L$ = 40, 20, and 10 mm, respectively. There was no significant difference in the macroscopic fracture patterns compared with the test results obtained under the same conditions. Therefore, only representative data are shown in Figs. 8-10. For the specimens with the shatterproof film, the film still held the glass fragments after fracture (Figs. 8(b), 9(b), and 10(b)). The fracture origin was confirmed at the points indicated by the arrows in the figures.

When $L$ = 40 mm, the fracture surface occurred in the direction parallel to the loading direction, starting from the glass surface on the opposite side of the point where the load was applied. This indicates that the fracture occurred from the location where the maximum tensile stress owing to bending occurred, i.e., bending failure occurred (Fig. 8). When $L$ = 10 mm, fracture surfaces occurred in a direction 45° to the loading direction from near the point where the load was applied (Fig. 10). This indicates that the shear stress generated by contact with the punch caused the shear failure, which is a Hertzian fracture in a two-dimensional contact problem. In two-dimensional disk contact problems based on the Hertzian contact theory [31], the maximum shear stress occurs in the direction of 45° to the contact axis [32].

When $L$ = 20 mm without the film, the fracture surface occurred in the direction parallel to the loading direction, as in the case of $L$=40 mm (Fig. 9(a)); however, with the film, a fracture surface with characteristics of both the $L$ = 40 and 10 mm cases appeared (Fig. 9(b)). Fracture origins were observed both near the point of load application and on the surface opposite the point of load application. Thus, at $L$ =20 mm, failure occurred in a mixed mode of bending and Hertzian fracture. Thus, by changing the support span of the three-point bending test, the failure mode of the glass changed, and bending failure, Hertzian fracture, and a mixed mode of bending and Hertzian fracture were observed.

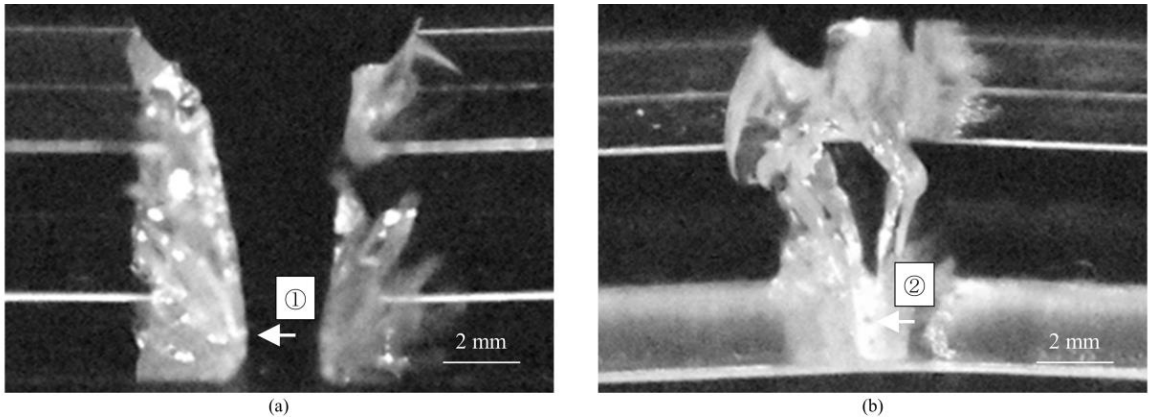

**Fig. 8** Vicinity of the fracture location of specimens ($L$ = 40 mm); (a) without film, (b) with film

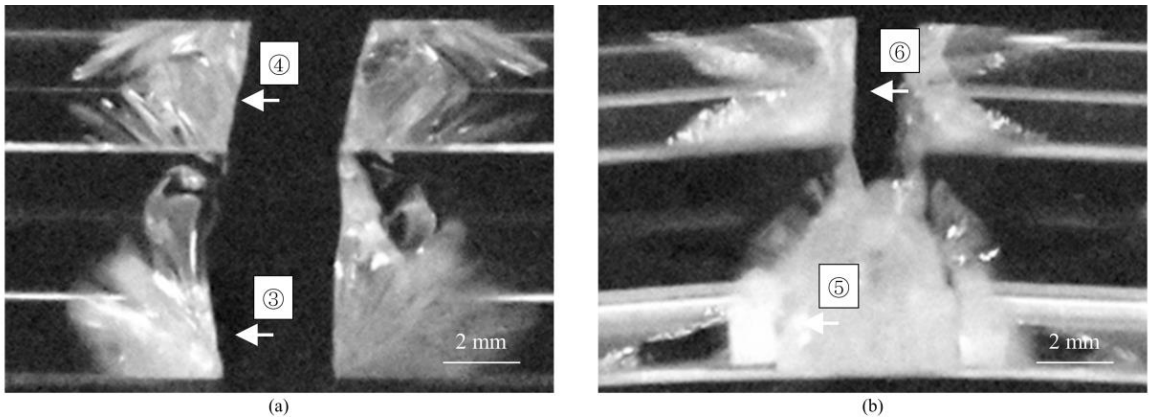

**Fig. 9** Vicinity of the fracture location of specimens ($L$ = 20 mm); (a) without film, (b) with film

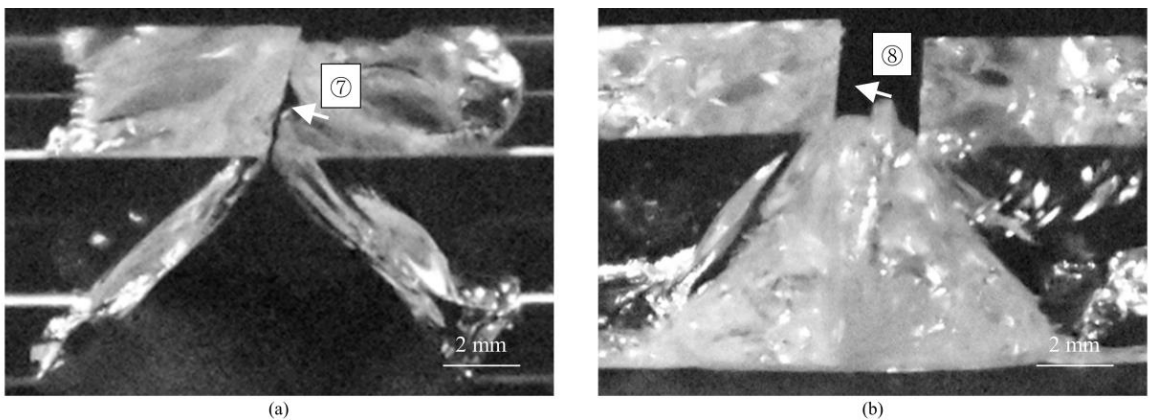

**Fig. 10** Vicinity of the fracture location of specimens ($L$ = 10 mm); (a) without film, (b) with film

Figures 11–13 show the fracture surfaces observed near the fracture origin of the specimens. The microscopic observation results were qualitatively similar to those obtained under the same conditions. Therefore, only representative data are shown in Figs. 11-14. In the case of $L$ = 40 mm, a mirror surface extending from the edge opposite the load application point was observed (Fig. 11), along with hackles that extended outwards from the mirror surface. From these fracture surfaces, it was confirmed that fracture occurred due to the maximum tensile stress caused by bending deformation. In the specimen without the film, the fracture origin was located at the point on the edge along the width of the most stressed cross-section on the opposite side of the loaded point (Fig. 11(a)), whereas, in the specimen with the film, the fracture origin was located at the corner edge of the most stressed cross section (Fig. 11(b)). It is possible that the applied film suppressed the opening of cracks originating from an in-plane Griffith flaw.

In the case of $L$ = 20 mm, as in the case of $L$ = 40 mm, a mirror surface and hackles were observed on the surface opposite the point of load application (Figs. 12(a) and (c)). In addition, Wallner lines were observed near the load-applied surface (Figs. 12(b) and (d)). This indicates that the crack propagated from both the loaded surface and the opposite surface, confirming the occurrence of a mixed mode of bending failure and Hertzian fracture.

In the case of $L$ = 10 mm, no fracture origin was observed on the surface opposite the load-applied surface. However, Wallner lines centred on the side of the load-applied surface were observed near the load-applied surface (Fig. 13). Therefore, only the Hertzian fracture occurred at $L$ =10 mm, and no bending failure was observed. As described above, when the support span $L$ is sufficiently long, only bending failure occurs, whereas as it shortens, the failure mode changes, and Hertzian fracture occurs. When the support span was sufficiently short, bending failure did not occur, and the specimen failed only by a Hertzian fracture. The characteristics of the failure modes were similar for the specimens with and without film.

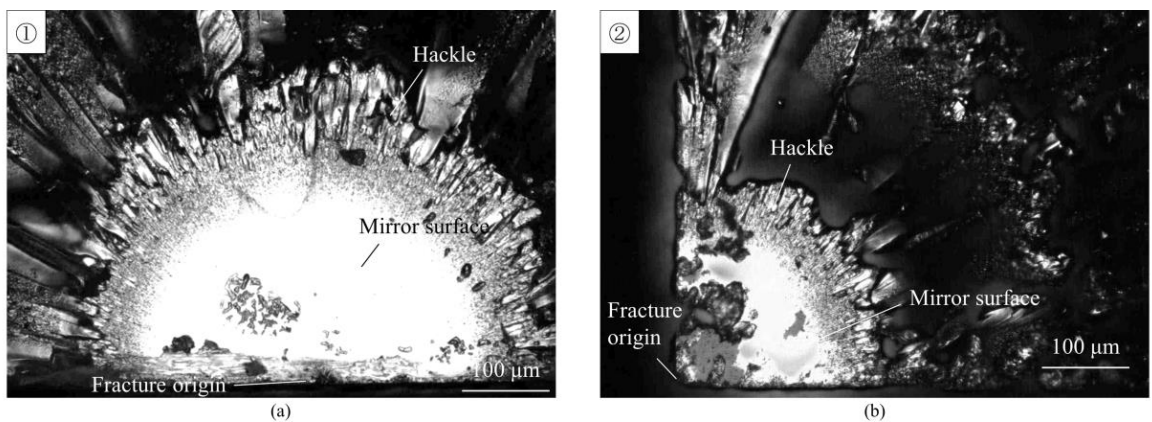

**Fig. 11** Fracture surface near fracture origin ($L$ = 40 mm); (a) without film, (b) with film

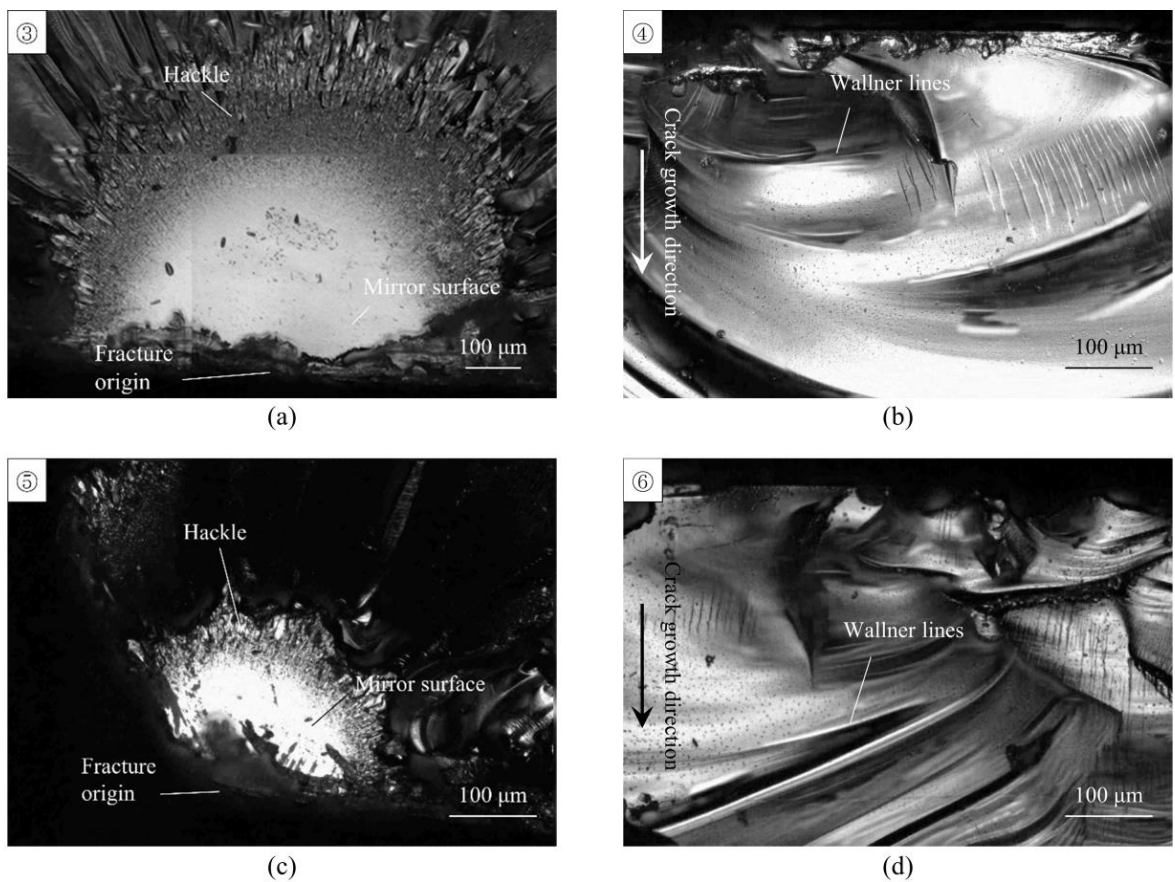

**Fig. 12** Fracture surface near fracture origins ($L = 20$ mm); (a) without film, on the opposite side of load application point, (b) without film, on side of load application point, (c) with film, on the opposite side of load application point, (d) with film, on side of load application point

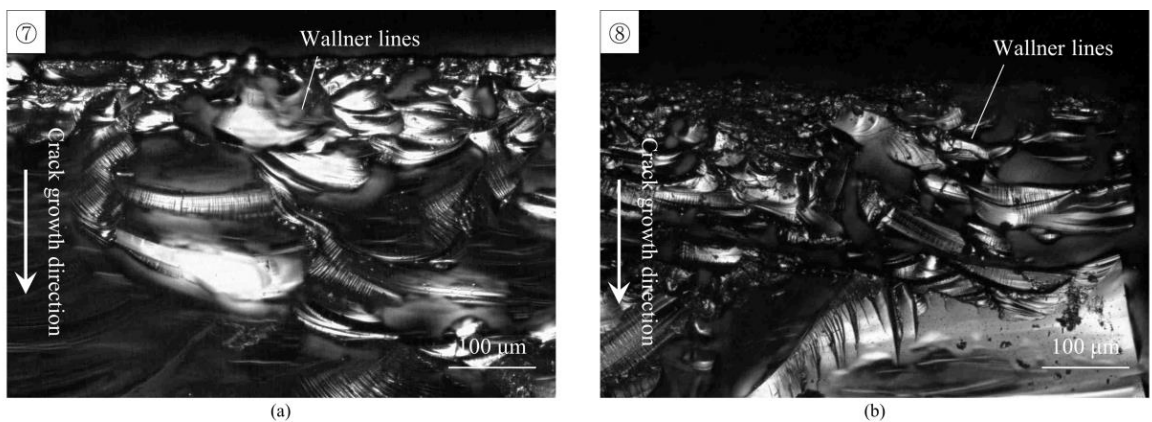

**Fig. 13** Fracture surface near fracture origin ($L = 10$ mm); (a) without film, (b) with film

Figure 14 shows the fracture surfaces at $L = 20$. As shown by the dashed lines in the figure, these fracture surfaces can be divided into three regions: the region where the crack propagates from the opposite side of the load-applied surface owing to bending failure, the region where the crack propagates from the load-applied surface owing to Hertzian fracture, and the intermediate region where these fracture surfaces are connected. From Fig. 14, the fractured area due to bending failure is smaller on the specimen with the film applied than on that without, while the area due to Hertzian fracture is larger when the film is applied. This may be due to the film suppressing crack opening in Mode I (opening mode) caused by the bending of the glass, which could improve its strength, although no significant difference in the breaking load was observed under the present conditions.

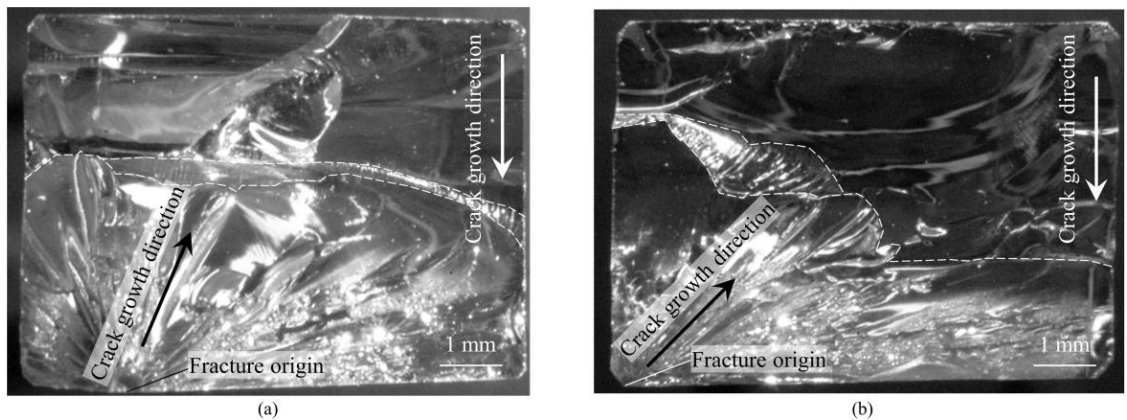

**Fig. 14** Fracture surface ($L$ = 20 mm); (a) without film, (b) with film

**4 Discussion**

To clarify the strength characteristics of glass with the shatterproof film under a bending load, we compared the experimental results of breaking load with the bending theory of a beam. Figure 15 shows the experimental values of the breaking load and theoretical critical load at which bending failure occurs, obtained using Eqs. (1)–(6). The theoretical values are the curves that increase as $L$ decreases. The theoretical curve for the composite beam (glass with film) had a slightly higher load than that for the beam without film. However, the two curves were almost identical. In the experimental results for $L = 40$ mm, where no Hertzian fracture occurred, the values of the breaking load with and without the film were almost identical and consistent with the theoretical value.

For $L = 40$ mm, the value of the experimental breaking load and the theoretical bending value were almost identical, confirming that bending failure was the dominant phenomenon. By contrast, for $L = 20$, the mean value of the experimental breaking load was larger than the theoretical value. For $L = 10$ mm, the difference between the two values was even larger. This may be because the stress field inside the glass was formed not only by bending deformation but also by contact with the punch and support. In the cases of $L = 20$ and 10 mm, the variation in the experimental results was larger than that in the case of $L = 40$ mm. As a result of a stress field formed by these two phenomena, this condition may become unstable. At $L = 10$ mm, the breaking load of the specimen without film was considered to be the critical load corresponding to the shear strength of glass (indicated by the chain line in Fig. 15) because no characteristic of bending failure was observed under this condition.

Thus, with the breaking load, the failure mode of the glass transitioned from bending failure to Hertzian fracture as the support span increased. However, under the present conditions, no significant difference was observed between the experimental values of the breaking load with and without the film.

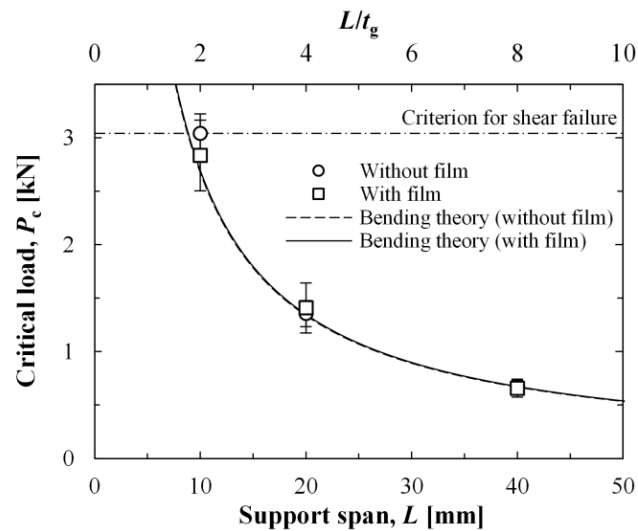

Fig. 15 Breaking loads and bending theory of a beam

Figure 16 shows the change in the theoretical critical load as a function of the thickness of the polymer layer applied to the glass. From Fig. 16, as the film-to-glass thickness ratio increased, the strength enhancement effect increased. This is because the film bears the stress through its deformation, thereby reducing the bending stress that occurs in the glass. However, under the present condition of thickness ratio of 3%, the strength enhancement effect was almost negligible. Future experiments with different thickness ratios are required to verify these results. In a previous study [19], the strength under dynamic loading improved at a thickness ratio similar to that of the present condition. Further studies are required to investigate the strength-enhancing effect of applying a polymer layer under dynamic loading.

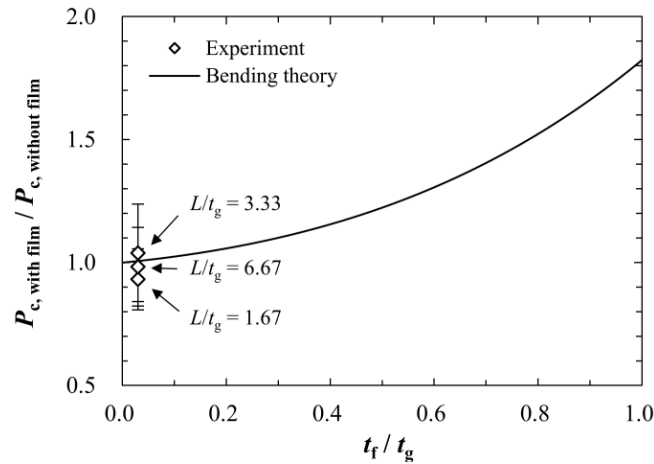

Fig. 16 Effect of film thickness on strength based on the bending theory of a composite beam

**5 Conclusion**

To clarify the effect of polymer film application on the failure mode and strength of a glass plate subject to quasi-static bending and indentation loads, three-point bending tests and fracture surface observations were conducted. By varying the support span of the specimen, the deformation mode was varied to generate three types of failures: bending failure, shear failure caused by Hertzian contact stress, and mixed-mode failure. Under the conditions in the present study, the breaking loads of the specimens with and without film were almost the same; however, the fracture surface observation indicated that the area subjected to a shear failure caused by Hertzian contact stress was larger on specimens with film application. Finally, the effect of film thickness on the breaking load due to bending deformation was theoretically predicted by determining the impact of the film-to-glass thickness ratio on strength improvement. Further experiments with different film thickness ratios are required to validate these results.

**Acknowledgments**

The authors appreciate Hideyasu Aoki and Katsumasa Aoki of SOUSHOW Co., Ltd. for their support of this research. This research did not receive any specific grant from funding agencies in the public, commercial, or not-for-profit sectors.

**References**

[1] A.A. Griffith, VI. The phenomena of rupture and flow in solids, Philos. Trans. R. Soc. London. Ser. A, Contain. Pap. a Math. or Phys. Character. 221 (1921) 163–198. https://doi.org/10.1098/rsta.1921.0006.

[2]     K. Osnes, O.S. Hopperstad, T. Børvik, Rate dependent fracture of monolithic and laminated glass: Experiments and simulations, Eng. Struct. 212 (2020) 110516. https://doi.org/10.1016/J.ENGSTRUCT.2020.110516.

[3]     Z. Lei, E. Rougier, E.E. Knight, M. Zang, A. Munjiza, Impact Fracture and Fragmentation of Glass via the 3D Combined Finite-Discrete Element Method, Appl. Sci. 11 (2021) 2484. https://doi.org/10.3390/APP11062484.

[4]     J. Rivera, J. Berjikian, R. Ravinder, H. Kodamana, S. Das, N. Bhatnagar, M. Bauchy, N.M.A. Krishnan, Glass Fracture Upon Ballistic Impact: New Insights From Peridynamics Simulations, Front. Mater. 6 (2019) 239. https://doi.org/10.3389/FMATS.2019.00239/BIBTEX.

[5]     K. Osnes, J.K. Holmen, O.S. Hopperstad, T. Børvik, Fracture and fragmentation of blast-loaded laminated glass: An experimental and numerical study, Int. J. Impact Eng. 132 (2019) 103334. https://doi.org/10.1016/J.IJIMPENG.2019.103334.

[6]     J. Ge, G.Q. Li, S.W. Chen, Theoretical and experimental investigation on fragment behavior of architectural glass panel under blast loading, Eng. Fail. Anal. 26 (2012) 293–303. https://doi.org/10.1016/j.engfailanal.2012.07.022.

[7]     H.D. Hidallana-Gamage, D.P. Thambiratnam, N.J. Perera, Failure analysis of laminated glass panels subjected to blast loads, Eng. Fail. Anal. 36 (2014) 14–29. https://doi.org/10.1016/j.engfailanal.2013.09.018.

[8]     S. Müller-Braun, C. Brokmann, J. Schneider, S. Kolling, Strength of the individual glasses of curved, annealed and laminated glass used in automotive windscreens, Eng. Fail. Anal. 123 (2021). https://doi.org/10.1016/j.engfailanal.2021.105281.

[9]     G. Deng, W. Ma, Y. Peng, S. Wang, S. Yao, S. Peng, Experimental study on laminated glass responses of high-speed trains subject to windblown sand particles loading, Constr. Build. Mater. 300 (2021) 124332. https://doi.org/10.1016/J.CONBUILDMAT.2021.124332.

[10]    X.H. Huang, X. er Wang, J. Yang, Z. Pan, F. Wang, I. Azim, Nonlinear analytical study of structural laminated glass under hard body impact in the pre-crack stage, Thin-Walled Struct. 167 (2021) 108137. https://doi.org/10.1016/J.TWS.2021.108137.

[11]    A. Vedrtnam, S.J. Pawar, Laminated plate theories and fracture of laminated glass plate – A review, Eng. Fract. Mech. 186 (2017) 316–330. https://doi.org/10.1016/J.ENGFRACMECH.2017.10.020.

[12]    T. Binar, J. Švarc, P. Vyroubal, T. Kazda, S. Rolc, A. Dvořák, The comparison of numerical simulation of projectile interaction with transparent armour glass for buildings and vehicles, Eng. Fail. Anal. 92 (2018) 121–139. https://doi.org/10.1016/j.engfailanal.2018.05.009.

[13]    H.D. Hidallana-Gamage, D.P. Thambiratnam, N.J. Perera, Numerical modelling and analysis of the blast performance of laminated glass panels and the influence of material parameters, Eng. Fail. Anal. 45 (2014) 65–84. https://doi.org/10.1016/j.engfailanal.2014.06.013.

[14] X. Xu, S. Chen, D. Wang, M. Zang, An efficient solid-shell cohesive zone model for impact fracture analysis of laminated glass, Theor. Appl. Fract. Mech. 108 (2020) 102660. https://doi.org/10.1016/J.TAFMEC.2020.102660.

[15] S. Wang, Y. Peng, X. Chen, K. Wang, The crack propagation and dynamic impact responses of tempered laminated glass used in high-speed trains, Eng. Fail. Anal. 134 (2022). https://doi.org/10.1016/j.engfailanal.2021.106024.

[16] P. Dietz, Research into the Use of Polymeric Film to Enhance the Safety of Sideglass, SAE Trans. 110 (2001) 782–785.

[17] S. Van Dam, J. Pelfrene, S. De Pauw, W. Van Paepegem, Experimental study on the dynamic behaviour of glass fitted with safety window film with a small-scale drop weight set-up, Int. J. Impact Eng. 73 (2014) 101–111. https://doi.org/10.1016/J.IJIMPENG.2014.06.002.

[18] T. Kojima, M. Notomi, T. Tsuji, Puncture Impact Test on Glass Plate Fitted with Shatterproof Window Film, Zair. Soc. Mater. Sci. Japan. 70 (2021) 796–801. https://doi.org/10.2472/JSMS.70.796.

[19] T. Kojima, K. Utakawa, M. Notomi, Improving the Impact Resistance of Glass Plate by Sticking Thin Film, J. Jpn. Soc. Exp. Mech. 18 (2018) 103–109. https://doi.org/10.11395/JJSEM.18.103.

[20] T. Kojima, Puncture Impact Behavior of Glass Plate Fitted With a Polymeric Film, Proc. ASME Int. Mech. Eng. Congr. Expo. 12 (2021). https://doi.org/10.1115/IMECE2020-23662.

[21] T. Kojima, K. Utakawa, M. Notomi, The Fracture Mode of a Glass Plate Stuck with Thin Film under a low-velocity Impact, in: 41st Solid Mech. Conf. B. Abstr., IPPT PAN, Warszawa, 2018, pp. 40–41.

[22] N. Shinkai, The Fracture and Fractography of Flat Glass, in: Fractography Glas., Springer US, Boston, MA, 1994: pp. 253–297. https://doi.org/10.1007/978-1-4899-1325-8_8.

[23] T. Kojima, M. Suzuki, Y. Tajiri, K. Utakawa, M. Notomi, Fracture mode of glass plate subject to low-velocity impact: experimental investigation and finite element simulation, Mech. Eng. J. 6 (2019) 19-00316. https://doi.org/10.1299/mej.19-00316.

[24] International Organization for Standardization, ISO 14704:2016 - Fine ceramics (advanced ceramics, advanced technical ceramics) — Test method for flexural strength of monolithic ceramics at room temperature, International Organization for Standardization, Geneva, 2016.

[25] D. Zenkert, An Introduction to Sandwich Structures, (1995). http://urn.kb.se/resolve?urn=urn:nbn:se:kth:diva-263054 (accessed November 29, 2022).

[26] H. Hertz, On the Contact of Elastic Solids, in: Misc. Pap., Macmillan and Co., New York, 1896, pp. 146–162.

[27] K.L. Johnson, Contact Mechanics, Cambridge University Press, Cambridge, 1985.